# Spatially controlled formation of superparamagnetic (Mn,Ga)As nanocrystals in high temperature annealed (Ga,Mn)As/GaAs superlattices

J. Sadowski[1,2], J. Z. Domagala[2], R. Mathieu[3], A. Kovacs[4] and P. Dluzewski[2]

[1] MAX-IV laboratory, Lund University, P.O. Box 118, SE-221 00 Lund, Sweden

[2] Institute of Physics, Polish Academy of Sciences, al. Lotników 32/46, PL-02-668 Warszawa, Poland

[3] Department of Engineering Sciences, Uppsala University, P.O. Box 534, SE-751 21 Uppsala, Sweden

[4] Ernst Ruska-Centre and Peter Grünberg Institute, Research Centre Jülich, 52425 Jülich, Germany

The annealing-induced formation of (Mn,Ga)As nanocrystals in (Ga,Mn)As/GaAs superlattices was studied by X-ray diffraction, transmission electron microscopy and magnetometry. The superlattice structures with 50Å thick (Ga,Mn)As layers separated by 25, 50 and 100Å thick GaAs spacers were grown by molecular beam epitaxy at low temperature (250 $^{o}$C), and then annealed at high temperatures of: 400, 560 and 630 $^{o}$C. The high temperature annealing causes decomposition to GaMnAs ternary alloy and formation of (Mn,Ga)As nanocrystals inside the GaAs matrix. The nanocrystals are confined in the planes that were formerly occupied by (Ga,Mn)As layers for up to the 560°C of annealing and diffuse throughout the GaAs spacer layers at 630°C annealing. The corresponding magnetization measurements show the evolution of the magnetic properties of as-grown and annealed samples from ferromagnetic, through superparamagnetic to the combination of both.

Since the very beginning of the research activity devoted to GaMnAs ferromagnetic semiconductor,[1-3] beside single phase ternary alloy with Mn ions partially replacing Ga in GaAs, also the nonhomogenous material comprising MnAs nanoinclusions embedded in GaAs lattice has been studied.[4] The metastable character of GaMnAs ternary alloy makes it fairly easy to produce an ensemble of metallic (Mn,Ga)As nanocrystals embedded in GaAs semiconductor host, just by high temperature (HT) post-growth annealing of homogeneous GaMnAs.[4-8] Similar effect can be achieved by implantation of Mn ions into GaAs,[9] however, this process introduces a lot of defects into GaAs matrix, so the former procedure is much more reliable. Even though the phase segregation and formation of metallic (Mn,Ga)As nanocrystals upon HT annealing of GaMnAs is quite easy to achieve, the control over dimensions, structures, densities and distribution of these nanocrystals in the GaAs matrix is not trivial. Binary MnAs is a well known metallic ferromagnet occurring in NiAs-type hexagonal structure in the bulk form. It has both structural (hexagonal-to-orthorhombic) and paramagnetic-to-ferromagnetic phase transition at the critical temperature (Tc) close to 40$^{o}$C.[10] MnAs layers can be epitaxially grown on GaAs substrates with thoroughly controlled growth conditions and parameters.[11] (Mn,Ga)As nanocrystals in GaAs host can form in a cubic (zinc-blende-structure) phase having a critical size of about 10 nm,[6-8,12-15] or in hexagonal phase typical for MnAs bulk, when their sizes are larger. Since the composition of these nanocrystals is not sufficiently well known (it is likely that there is some admixture of gallium in them) here we are using a notation of (Mn,Ga)As due to the possible Ga content in the nanocrystals.[6-8,13] So far it is relatively well known how to control the crystalline structure of (Mn,Ga)As nanocrystals in HT annealed GaMnAs e.g. low annealing temperatures (500 – 560 $^{o}$C) yield cubic nanocrystals whereas high annealing temperatures (600 - 650 $^{o}$C) result in hexagonal nanocrystal formation.[6-8,12-15] In contrast, the control of their distribution has not been thoroughly investigated. Here we examine the possibility of confining (Mn,Ga)As nanocrystals in-plane by using short-period GaMnAs/GaAs superlattices as the primary samples from which the nanoinclusions are generated upon HT post-growth annealing procedures. The structures with MnAs nanocystals fabricated by HT annealing of superlattices containing GaMnAs have already been studied by Shimizu and Tanaka in 2001.[16] In their report the GaMnAs/AlAs SLs with 5 nm thick AlAs spacers and 5 to 50 nm thick GaMnAs layers have been investigated, however only optical properties of such structures were reported in Ref. 16 and no detailed investigations concerning nanocrystals' structure and distribution were performed.

We have studied three GaMnAs/GaAs superlattices with Mn content of 3.7% in GaMnAs layers, and thicknesses of GaMnAs equal to 18 molecular layers (ML), i.e. about 50 Å; and three different thicknesses of GaAs spacer layers of 9, 18 and 36 ML; or 25, 50 and 100Å – SL-1, 2, and 3, respectively. The number of repetitions of the GaMnAs-GaAs sequence was 100 for each sample. The samples were grown by molecular beam epitaxy (MBE) in SVTA III-V MBE system. Arsenic was delivered from the valve cracker source, with cracking zone operated at 950 $^{o}$C, i.e. As$_2$ flux was used for GaMnAs growth. The SL structures were grown at low substrate temperature (about 250 $^{o}$C) with the As$_2$/Ga flux ratio of about 1.5. The samples were glued on molybdenum holders by liquid In, which provides good thermal contact and lateral temperature uniformity of the substrates during the MBE growth. After the MBE growth the samples were taken out of the MBE system, and cleaved into 4 parts. One part was left intact; three other parts were mounted again on the substrate holders and reintroduced into the MBE system for HT annealing. The annealing temperatures were chosen to be 400, 560 and 630 $^{o}$C. At each temperature the pieces of three different SL structures were annealed simultaneously, i.e. they were placed on the same holder. The annealing temperature was measured by the MBE substrate heater thermocouple, with temperature calibrated by the surface reconstruction transitions of the GaAs(001).

Fig. 1 shows XRD curves detected in 2θ/ω mode, so called 2θ/ω scans, around 004 Bragg reflections of all three SL as-grown and annealed samples; starting with the SL with the shortest period (SL-1). Figs 1a, 1c and 1e show broader scans (2θ range changes up to $6^0$) with clearly visible satellite peaks associated with the superlattice periodicity. Figs 1b, 1d and 1f show narrower region close to the 0-order SL peaks, indicated by vertical arrows. All three SLs show up to 2-nd order satellite diffraction peaks. The absence of higher order satellites may be due to the low chemical contrast and small lattice parameter difference between GaMnAs layers and GaAs spacers (about 0.1%).

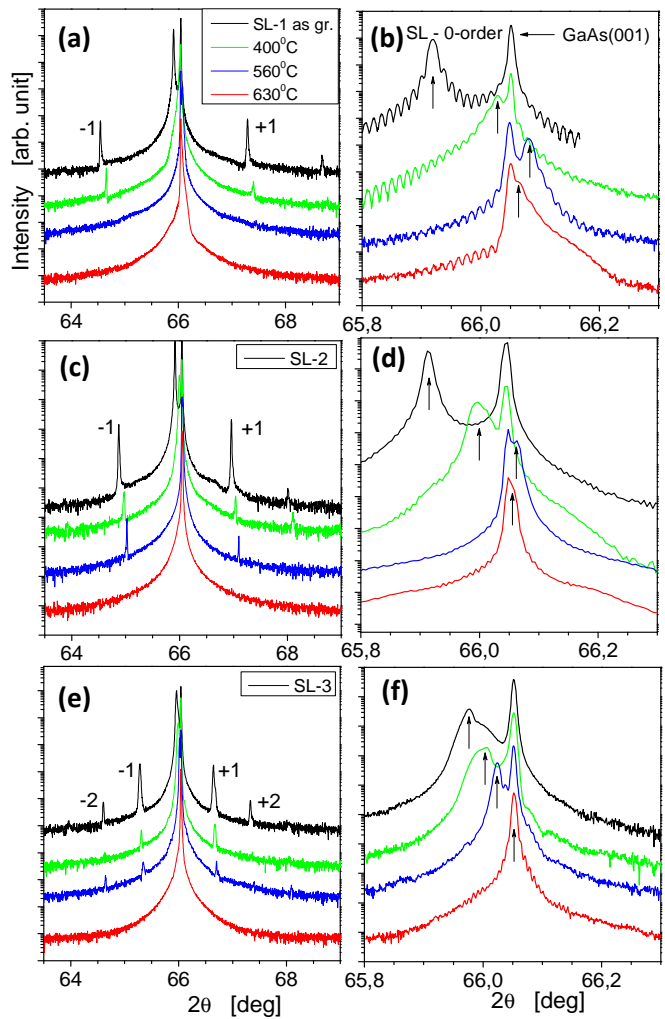

FIG. 1. (color online) 2θ/ω scans around 004 Bragg peak for three GaMnAs/GaAs superlattices with 50 Å thick GaMnAs layers and 25, 50 and 100 Å thick GaAs spacers – SL-1, 2 and 3, respectively. In each panel the upper curves are taken for the as-grown samples, lower curves - for samples annealed

at 400, 560 and 630 °C, consecutively. Panels (b), (d), (f) show annealing induced changes of the angular positions of the 0-order SL diffraction peaks (indicated by vertical arrows) reflecting changes of the strain in the entire SL structures due to the HT annealing. Position of the GaAs(001) substrate peak is not dependent on the annealing and is equal to 66.045$^0$.

As can be seen in Fig. 1, the post-growth annealing considerably reduces the intensities of satellite peaks for all SLs structures; in particular, annealing at the highest temperatures (630 °C) smears them out completely. The decrease of the satellite peak intensities may be due to the partial diffusion of the MnAs nanocrystals into the GaAs spacer layers. Since after the 630 °C annealing no satellite peaks are visible we can conclude that upon this highest annealing the (Mn,Ga)As nanocrystals completely diffuse even throughout the thickest spacer which is 100 Å for SL-3. For intermediate annealing temperature i.e. 560 °C the satellites disappear only for SL-1 which has the thinnest GaAs spacer of 25 Å, and are still present for other SL structures with 50Å thick and 100 Å thick spacers (samples SL-2 and SL-3, respectively). By comparing the 560 °C annealing for SL-1 and SL-2 samples, with 25 Å and 50 Å thick GaAs spacers, respectively, we can estimate the lower limit of the diffusion constant of Mn at this temperature. Applying the Fick's diffusion law, the profile of Mn concentration (n) in spacer layers can be described by:

$$n(x,t) = n(0) erfc\left(\frac{x}{2\sqrt{Dt}}\right) \quad (1)$$

where *erfc* is the error function, *n* is the Mn concentration, *D* - diffusion constant, *x* - a spatial coordinate in the growth direction (i.e. perpendicular to the SL planes) and *t* is time. Knowing that Mn in SL-1 structure is completely interdiffused throughout 25 Å thick spacer at the 560 °C annealing, we can estimate the lower limit of the value of D. If the satellites of XRD peaks disappear, then Mn should diffuse at least to the half of the spacer distance which is about 13 Å in the case of SL-1 (we can assume that Mn can diffuse in both directions, i.e. towards the substrate and towards the surface). The complete interdiffusion means that the concentration of Mn in the spacers is half of the initial Mn concentration in the GaMnAs SL layers after the annealing time of 0.5h. Entering these parameters into formula (1) we get the value of $D \sim 10^{-21} m^2/s$. This is one order of magnitude higher than D value estimated by us previously[17] for diffusion of Mn interstitials across LT GaAs, however during annealing at much lower temperature (200 $^o$C) than applied for HT annealing experiments reported here. It is known[18], that the binding energy of Mn interstitials in GaMnAs is much lower than that of Mn in Ga sites; however our results presented here (and published elsewhere[15]) show that diffusion of Mn from Ga sites takes place already at the temperatures close to 400 $^o$C, which is only about 130 $^o$C higher than the temperature at which GaMnAs with low Mn content (1 % and below) can be grown.[19] This can easily be seen in XRD results shown in the left panels of Fig.1. The intensities of satellites of SL peak shown in Fig.1a are considerably reduced after annealing at 400 $^o$C, moreover, as shown in the right panels of Fig.1, the angular position of the zero order (main) SL diffraction peak moves significantly towards that of GaAs(001) substrate, which indicates that GaMnAs ternary alloy partially decomposes already at 400 $^o$C. Similarly to the case of HT annealed single GaMnAs layers the angular position of zero order SL diffraction peaks reflecting the averaged strain of the whole SL structure moves to the higher diffraction angles at 560 $^o$C annealing and then moves slightly back to the lower diffraction angles at 630 $^o$C annealing. This has been verified previously by us[15,20] and by other groups[6,7,12] as a strain effect of (Mn,Ga)As nanocrystals on surrounding GaAs lattice. Cubic nanocrystals prevailing at lower annealing temperarure (560 $^o$C) exert higher strain on GaAs, than hexagonal nanocrystals formed at 630 $^o$C annealing.[15,21] Interestingly SL-3 with the thickest GaAs spacers does not follow this trend and exhibits also different temperature dependence of magnetization (see below). The different dependence of strain on annealing temperature for SL-3 may be due to the contribution of spacer layers, which contain HT annealed low-temperature GaAs, where pure As nanocrystals are generated during the HT post-growth annealing.[22] These As nanocrystals also exert strain on the surrounding GaAs, in SL-3 sample with LT GaAs layers 2 times thicker than GaMnAs layers, these effects can be prevailing.

The conclusions drawn from the XRD results are further confirmed by the scanning transmission electron microscopy (STEM) investigations of SL-2 with 50Å thick GaAs spacer layers. Fig. 2 shows cross-sectional annular-dark field (ADF) STEM images of SL-2 annealed at 560 ° and 630 °C – panels (a) and (b), respectively.

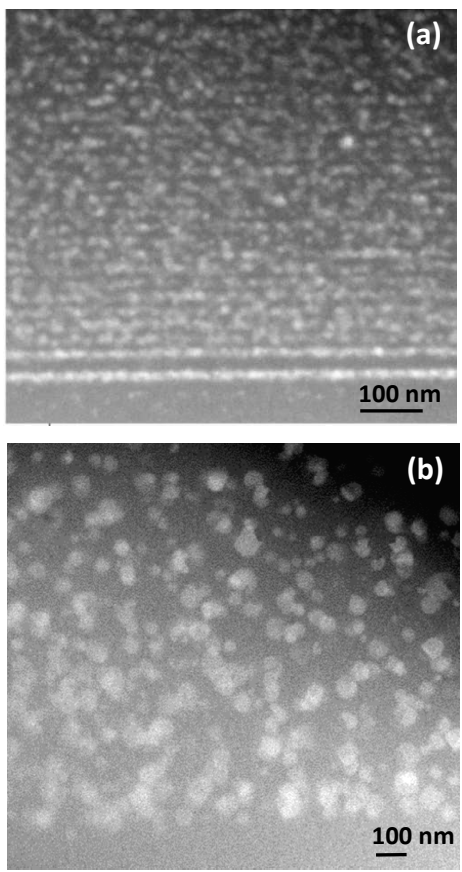

FIG. 2. Cross-sectional STEM images of the region close to the GaAs(001) substrate of SL-2 annealed at: (a) 560 °C, the in-plane correlation of nanocrystals is noticeable; (b) 630 °C, the nanocrystals are much larger in comparison to those formed at 560 °C annealing, no more in-plane correlation of nanocrystals is visible.

STEM images were acquired for pieces of SL-2 annealed at 560 °C and 630 °C. It can be seen in Fig.2a, that after 560 °C annealing the (Mn,Ga)As nanocrystals are correlated and preferentially located in the planes formerly occupied by GaMnAs. After 630 °C annealing the size of the nanocrystals increases significantly and their in-plane correlation is lost i.e. they completely diffuse across the LT GaAs spacers, as shown in Fig. 2 (b). These results confirm our earlier investigations of the HT annealing of single GaMnAs layers inside TEM, showing that upon HT annealing the emerging nanocrystals move, and grow in size due to their coalescence.[14] Interestingly the preferential location of the nancorystals at regions formerly occupied by GaMnAs layers is less pronounced in the direction towards the sample surface which may be due to the accumulation of Mn interstitials in this direction during the growth of the sample. Similar profile of Mn interstitials, i.e. increase of their concentration towards the sample surface has been observed previously for single layers,[23] but, to our knowledge it has not been reported for GaMnAs/GaAs superlattice structures.

The magnetization of the superlattices was recorded on a SQUID magnetometer from Quantum Design Inc. The magnetic field employed to probe the magnetization was applied along the -110 direction in all measurements.

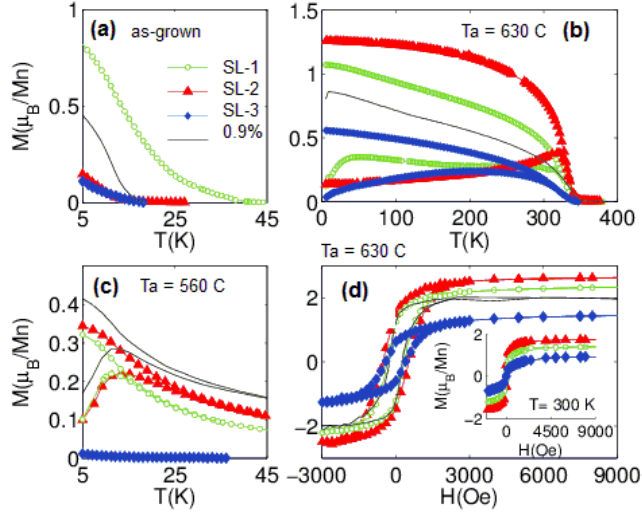

FIG. 3. (Color online) Temperature dependence and magnetic field dependence of magnetization measured in SQUID magnetometer for: (a) - as-grown, (b) 630 °C annealed and (c) 560 °C annealed SL-1, SL-2, SL-3 and single $Ga_{0.991}Mn_{0.009}As$ layer (curve labeled 0.9%). The M(T) curves were measured with external magnetic field of 20 Oe (a,b) and 50 Oe (c). The hysteresis loops shown in (d) are measured at 5K (main frame) and 300K (the inset) for 630 °C annealed samples.

Fig 3. shows the temperature dependence of magnetization (a-c) and hysteresis loops (d) for as-grown, 630 °C annealed and 560 °C annealed SLs and thick, single $Ga_{0.991}Mn_{0.009}As$ layer. The as-grown SLs show some magnetic ordering below 40K for SL-1, and below 20K for SL-2 and SL-3 (see Fig.3a). A thick, single $Ga_{0.991}Mn_{0.009}As$ layer, labeled "0.9%" was found to order magnetically below 20K as well[15] (although in Ref. 15 this layer is labeled by its nominal composition of 0.5%) and its magnetization is added for comparison. The effect of the annealing on the magnetism of the superlattices is indeed similar to the single layer case. Akin to the bulk low-doped (Ga,Mn)As single layers studied in Ref. 15 no significant magnetic signal was observed after annealing the SLs at 400 °C. However a superparamagnetic-like signal was observed after annealing at 560 °C, with a maximum in the zero-field-cooled magnetization below T=15K (see Fig.3c). The signal is much weaker for SL-3, however we have verified previously, that the FM phase transition temperature is reduced, or eventually the transition disappears, in the short period GaMnAs/GaAs superlattices with thick GaAs spacer layers.[24] As shown in Fig.3b, the 630 °C annealed SLs exhibit a ferromagnetic-like ordering below 350K, again similarly to the single layers studied earlier in the same conditions.[15] Similar high temperature ferromagnetism was observed in (Ga,Mn)As systems containing (Mn,Ga)As nanocrystals with zinc-blende structure[12] as well as, interestingly, in strained single epitaxial layers of MnAs deposited on GaAs (110)[25] and GaAs(100).[26] In situ high resolution TEM annealing studies have revealed the formation of cubic ferromagnetic nanocrystals in low Mn-doped single layers such as $Ga_{0.991}Mn_{0.009}As$.[14]

The similarity in the magnetic properties of the superlattices and low Mn content single GaMnAs layers after HT annealing is also evidenced in the magnetic field dependent measurements. Hysteresis loops were recorded at 5K (Fig.3d, the main frame) and room temperature (Fig3d, the inset). The low-temperature coercivities of the SLs are similar to those of the single layers and are between 400 and 500 Oe. At room temperature the coercivities amount to about 50 Oe. The saturation magnetization of the SLs reaches about 2.6 $\mu_B$/Mn at low temperature, and decreases to 1.7 $\mu_B$/Mn for SL-2 (see Fig.3d). Although it is interesting to notice that the saturation magnetization of different SLs varies significantly, it is difficult to analyze further the value of the associated magnetic moments, since the Mn-rich phases with different sizes and structures may co-exist in the layers. These phases have different magnetizations[27] and possibly different magnetic anisotropies.

In summary – we have investigated the formation and diffusion of (Mn,Ga)As nanocrystals fabricated by high temperature post growth annealing of short period GaMnAs/GaAs superlattices. XRD and STEM results show that annealing at 560 °C produces (Mn,Ga)As nanocrystals in the regions previously occupied by GaMnAs layers, which are confined in-plane, in SLs structure with 50 Å and 100 Å thick GaAs spacer layers. For the 25Å thick spacers the complete interdiffusion of (Mn,Ga)As nanocrystals through GaAs layers occurs. The diffusion coefficient of Mn atoms from Ga sites at 560°C, was determined to be at least $10^{-21} m^2/s$. Annealing at higher temperatures (630 °C) cause complete interdiffusion of (Mn,Ga)As nanocrystals throughout the GaAs spacers in all the superlattice structures studied. Measurements of magnetic properties indicate that the (Mn,Ga)As nanocrystals obtained by the high temperature post growth annealing of GaMnAs/GaAs superlattices have properties similar to those of the same nanocrystals occurring in HT annealed single GaMnAs layers (i.e. superparamagnetic or ferromagnetic for annealing at 560 °C and higher).

This work was partly supported by the "FunDMS" Advanced Grant of the European Research Council within the "Ideas" 7th Framework Programme of the European Commission and by the European Regional Development Fund within the Innovative Economy Operational Programme 2007-2013 No POIG.02.01-00-14-032/08. The Swedish Research Council (VR) and the Göran Gustafsson Foundation (Sweden) are also acknowledged for financial support of the MBE system at MAX-IV laboratory and the SQUID magnetometers at Uppsala University.

[1] H. Ohno, A. Shen, F. Matsukura, A. Oiwa, A. Endo, S. Katsumoto, and Y. Iye, Appl. Phys. Lett. **69**, 363 (1996).

[2] A. Van Esch, L. Van Bockstal, J. De Boeck, G. Verbanck, A. S. van Steenbergen, P. J. Wellmann, B. Grietens, R. Bogaerts, F. Herlach, and G. Borghs, Phys. Rev. B 56, 13103 (1997).

[3] J. Sadowski, J. Domagała, J. Bąk-Misiuk, K. Świątek, J. Kanski, L. Ilver, H. Oscarsson, Acta Phys. Pol. A, **94**, 509 (1998).

[4] J. De Boeck, R. Oesterholt, A. Van Esch, H. Bender, C. Bruynseraede, C. Van Hoof, and G. Borghs, Appl. Phys. Lett. 68, 2744 (1996).

[5] H. Akinaga, S. Miyanishi, K. Tanaka, W. Van Roy, and K. Onodera, Appl. Phys. Lett. **76**, 97 (2000).

[6] M. Moreno, A. Trampert, B. Jenichen, L. Däweritz, and K. H. Ploog, J. Appl. Phys. **92**, 4672 (2002).

[7] M. Moreno, B. Jenichen, L. Däweritz, and K. H. Ploog, J. Vac. Sci. Technol. B **23**, 1700 (2005).

[8] A. Kwiatkowski, D. Wasik, M. Kamińska, R. Bożek, J. Szczytko, A. Twardowski, J. Borysiuk, J. Sadowski, and J. Gosk, J. Appl. Phys. **101**, 113912 (2007).

[9] P. J. Wellmann, J. M. Garcia, J.-L. Feng, and P. M. Petroff, Appl. Phys. Lett. 71, 2532 (1997).

[10] J. B. Goodenough and J. A. Kafalas, Phys. Rev. 157, 389 (1967).

[11] M. Tanaka, J. P. Harbison, T. Sands, T. L. Cheeks, V. G. Keramidas, and G. M. Rothberg J. Vac. Sci. Technol. B 12, 1091 (1994).

[12] M. Yokoyama, H. Yamaguchi, T. Ogawa, and M. Tanaka, J. Appl. Phys. **97**, 10D317 (2005).

[13] K. Lawniczak-Jablonska, J. Libera, A. Wolska, M. T. Klepka, P. Dłużewski, J. Sadowski, D. Wasik, A. Twardowski, A. Kwiatkowski, and K. Sato, Phys. Status Solidi RRL **5**, 62 (2011).

[14] A. Kovács, J. Sadowski, T. Kasama, J. Domagała, R. Mathieu, T. Dietl and R. E. Dunin-Borkowski, J. Appl. Phys. **109**, 083546 (2011).

[15] J. Sadowski, J. Z. Domagala, R. Mathieu, A. Kovács, T. Kasama, R. E. Dunin-Borkowski and T. Dietl, Phys. Rev. B **84**, 245306 (2011).

[16] H. Shimizu and M. Tanaka, J. Appl. Phys. 89, 7281 (2001).

[17] J. Adell, I. Ulfat, L. Ilver, J. Sadowski, and J. Kanski, J. Phys. Cond. Mat. **23** 085003 (2011).

[18] K. W. Edmonds, P. Boguslawski, K. Y. Wang, R. P. Campion, S. N. Novikov, N. R. S. Farley, B. L. Gallagher, C. T. Foxon, M. Sawicki, T. Dietl, M. Buongiorno Nardelli and J. Bernholc Phys. Rev. Lett. **92**, 037201 (2004).

[19] H. Ohno, Science **281,** 951 (1998).

[20] J. Bak-Misiuk, J .Z .Domagala P. Romanowski, E. Dynowska, E. Lusakowska, A. Misiuk, W.Paszkowicz, J.Sadowski, A.Barcz, W.Caliebe, Radiation Physics and Chemistry **78**, S116 (2009).

[21] M. Moreno, V. M. Kaganer, B. Jenichen, A. Trampert, L. Däweritz, and K. H. Ploog, Phys. Rev. B **72**, 115206, (2005).

[22] M. Toufella, P. Puech, R. Carles, E. Bedel, C. Fontaine, A. Claverie, and G. Benassayag, J. Appl. Phys. 85, 2929 (1999).

[23] L. Horak, J. Matejova, X. Marti, V. Holy, V. Novak, Z. Soban, S. Mangold, and F. Jimenez-Villacorta, Phys. Rev. B **83**, 245209 (2011).

[24] J. Sadowski, R. Mathieu, P. Svedlindh, M. Karlsteen, J. Kanski, Y. Fu, J. Z. Domagala, W. Szuszkiewicz, B. Hennion, D. K. Maude, R. Airey, G.Hill, Thin Solid Films, **412**, 122 (2002).

[25] P. Xu, J. Lu, L. Chen, S. Yan, H. Meng, G. Pan, J. Zhao, Nanoscale Res. Lett. **6**, 125 (2011).

[26] J. M. Wikberg, R. Knut, S. Bhandary, I. di Marco, M. Ottosson, J. Sadowski, B. Sanyal, P. Palmgren, C. W. Tai, O. Eriksson, O. Karis, and P. Svedlindh, Phys. Rev. B **83**, 024417 (2011).

[27] S. Sanvito and N. Hill, Phys. Rev. B **62**, 15553 (2000).